\documentclass[10pt,aps,prb,twocolumn,showpacs,amssymb]{revtex4-1}
\usepackage{amsmath,graphicx,epsfig,mathrsfs}
\usepackage{bm}
\usepackage{epigraph}
\usepackage{geometry}    
\usepackage{ulem}

\usepackage{cases}
\usepackage{cancel}
\usepackage{float}
\usepackage{color}

\usepackage{color}
\definecolor{darkgreen}{rgb}{0.0, 0.5, 0.0}
\definecolor{brown}{rgb}{0.59, 0.29, 0.0}
\definecolor{darkgreen2}{rgb}{0.01, 0.75, 0.24}
\usepackage{hyperref}
\hypersetup{backref=true,
pagebackref=true,
hyperindex=true,
breaklinks=true,
urlcolor=cyan,
linkcolor=blue,
bookmarks=true,
colorlinks=true,
citecolor=blue,
bookmarksopen=true}

\begin{document}
\title{Tuning skyrmion Hall effect via engineering of spin-orbit interaction}
\author{Collins  Ashu  Akosa$^{1,2}$}
\email{collins.akosa@riken.jp}
\author{Hang Li$^{3}$}
\author{Gen Tatara$^{1,4}$}
\author{Oleg\,A.\,Tretiakov$^{5,6}$}
\email{o.tretiakov@unsw.edu.au}
\affiliation{$^{1}$RIKEN Center for Emergent Matter Science (CEMS), 2-1 Hirosawa, Wako, Saitama 351-0198, Japan.}
\affiliation{$^2$Department of Theoretical and Applied Physics, African University of Science and Technology (AUST), Km 10 Airport Road, 
Galadimawa, Abuja F.C.T, Nigeria.}
\affiliation{$^{3}$Institute for Computational Materials Science, School of Physics and Electronics, Henan University, Kaifeng 475004, China.}
\affiliation{$^4$RIKEN Cluster for Pioneering Research (CPR), 2-1 Hirosawa, Wako, Saitama, 351-0198 Japan.}
\affiliation{$^5$School of Physics, The University of New South Wales, Sydney 2052, Australia.}
\affiliation{$^6$National University of Science and Technology "MISiS", Moscow 119049, Russia.}

\date{\today}

\begin{abstract}
We demonstrate that the Magnus force acting on magnetic skyrmions can be efficiently tuned via modulation of the spin-orbit interaction strength. We show that the skyrmion Hall effect, which is a direct consequence of the non-vanishing Magnus force on the magnetic structure can be suppressed in certain limits. Our calculations show that the emergent magnetic fields in the presence of spin-orbit coupling (SOC) renormalize the Lorentz force on itinerant electrons and thus influence the topological transport. In particular, we show that for a N\'eel-type skyrmion and Bloch-type antiskyrmion, the skyrmion Hall effect (SkHE) can vanish by tuning appropriately the strength of Rashba and Dresselhaus SOCs, respectively. Our results open up alternative directions to explore in a bid to overcome the parasitic and undesirable SkHE for spintronic applications. 
\end{abstract}
\maketitle

\section{Introduction}
Over the last decade, spintronic research interest has switched towards a novel direction called 
\textit{spin-orbitronics} that exploits the relativistic coupling of electron's spin to its orbit to 
create new and intriguing effects and materials. \cite{Manchon2014,Kuschel2015, Chen2017} 
It turns out that the spin-orbit interaction is very crucial in effects related to the efficient conversion of charge to spin current, 
\cite{Sinova2015, Hoffmann2015} that is essential for spintronic applications. The former has been largely exploited for the creation of a novel class 
of topological materials such as chiral domain walls and magnetic skyrmions with enhanced thermal stability, low critical currents and smaller sizes. 
Therefore, spin-orbit related effects open up promising directions to create, manipulate, and detect spin currents for spintronic applications. 

In magnetic materials with broken inversion symmetry, an atom with strong SOC can mediate an antisymmetric exchange interaction called the 
Dzyaloshinskii-Moriya iteraction (DMI), that favors the non-collinear alignments of atomic spins. \cite{Dzyaloshinsky1958, Moriya1960} In such materials, 
the competition between the DMI and other magnetic interactions notably, the exchange (which favors collinear alignment of atomic spins) is essential for 
the stabilization of exotic magnetic states such as helimagnets \cite{Kishine2015} and magnetic skyrmions. \cite{Robler2006} The later have been widely 
projected as a viable contender for information carriers in the next-generation data storage and spin logic devices due to their small spatial extent, 
high topological protection, and efficient current-induced manipulation that allows for robust, energy-efficient, and ultra-high density spintronic applications. 
\cite{Nagaosa2013, Fert2013} However, the integration of ferromagnetic skyrmions in such applications is hindered by the undesirable SkHE, a transverse motion
 to the direction of current flow. \cite{Litzius2017, Jiang2017} 
 
 To overcome this parasitic effect, several proposals have been put forward such as, the use of 
 antiferromagnetic skyrmions, \cite{Barker2016, Zhang2016a, Gobel2017c, Akosa2018} edge repulsion, \cite{Zhang2017} magnetic bilayer-skyrmion, 
 \cite{Zhang2016} skyrmionium, \cite{Finazzi2013, Zheng2017,Zhang2018, Beg2015, Beg2017,Zhang2016c} antiskyrmions, \cite{Huang2017, Potkina2019} and via spin current partially polarized along the direction of applied current. \cite{Gobel2019b} These proposals focus on suppressing the inherent Magnus force in these systems, while little attention has been 
 paid to understanding its source, i.e., the nature of the texture-induced emergent magnetic field. Moreover, since the stabilization of topological magnetic textures such as skyrmions, usually requires strong SOC, it is important to investigate the effect of the latter on these fictitious electromagnetic fields.  Indeed, recent studies showed that SOC induces additional fictitious electric field that manifests itself as spin-motive force \cite{Kim2012, Tatara2013} on the itinerant electrons and gives raise to charge current \cite{Tatara2007} and chiral damping. \cite{Akosa2018b} However, the effect of the SOC-induced emergent magnetic field on the conduction electrons has more or less been overlooked. 

In this paper, we provide a theoretical framework that takes into account these fictitious magnetic fields and elucidate their impact on the topological transport inherent to magnetic skyrmions. We demonstrate that the SkHE, which is a direct consequence of a non-vanishing Magnus force on the magnetic structure can be efficiently tuned via modulation of the spin-orbit interaction strength. In particular, we show that for a N\'eel-type skyrmion and Bloch-type antiskyrmion, the SkHE can be tuned to zero via the modulation of the strength of the Rashba SOC (RSOC) and Dresselhaus (DSOC), respectively. Our results opens up alternative directions to overcome the parasitic and undesirable SkHE in ferromagnetic skyrmions.

\section{Theoretical Model}
It is known that DSOC stabilizes Bloch-type skyrmions, \cite{Muhlbauer2009,Yu2010, Seki2012} while RSOC stabilizes N\'eel-type skyrmions. 
\cite{Kezsmarki2015,Kurumaji2017} However, recent realization of Bloch-type skyrmions in Rashba metals \cite{Hayami2018,Rowland2016} motivates us here to consider an interplay of both types of SOC in a skyrmionic system described by the Hamiltonian
\begin{equation}\label{eq:ham0}
\hat{\mathcal{H}} = \frac{\hat{\bm p}^2}{2m^*}  +  J{\bm m}({\bm r},t)\cdot \hat{\boldsymbol{\sigma}}  + \hat{\mathcal{H}}_{ R} + \hat{\mathcal{H}}_{ D} , 
\end{equation} 
where $m^*$ is the effective mass of electrons, $\hat{\bm p}$ is the momentum operator, $J$ is the constant of exchange interaction between the local moments in the direction 
of ${\bm m}$ and spins of itinerant electrons described by vector of Pauli matrices $\hat{\boldsymbol{\sigma}}$. The terms
$\mathcal{H}_{ R} =  \beta_{ R}(\sigma_y p_x   -  \sigma_xp_y )/\hbar$ and $\mathcal{H}_{ D} =  \beta_{ D}( \sigma_xp_x  - \sigma_y p_y)/\hbar$ describe the RSOC and 
DSOC, respectively. To render our analysis trackable, we consider an isolated skyrmion (antiskyrmion) with analytical ansatz without loss of generality given in spherical 
coordinates as ${\bm m}({\bm r},t) = \big(\cos\Phi\sin\theta, \sin\Phi\sin\theta , \cos\theta \big)$, where the azimuthal and 
radial angles are given as $\Phi(x,y) = q\,\rm{Arg}\, (x,y) + c\pi/2$ and $\theta = \pi(3-p)/2 + 4 \tan^{-1} (e^{r/r_s})$, respectively, where $r_s$ is the skyrmion radius. In this 
representation, $p = \pm 1$ is the \textit{polarity} that defines the orientation of the skyrmion's core, $q=\pm1$ is the \textit{vorticity}, i.e., $q = +1$ for \textit{skyrmions} 
and $q=-1$ for \textit{antiskyrmions}, and $c$ is the \textit{heliticity}, such that $c=0$ for N\'eel-type and $c= \pm 1$ for Bloch-type skyrmions (antiskyrmions). Note that the topological charge $Q$ of the magnetic solitons is given by $Q=pq$. \cite{Tretiakov2007} 

\section{Analytical Results}
\begin{figure*}[t!]
\centerline{\includegraphics[width=12cm]{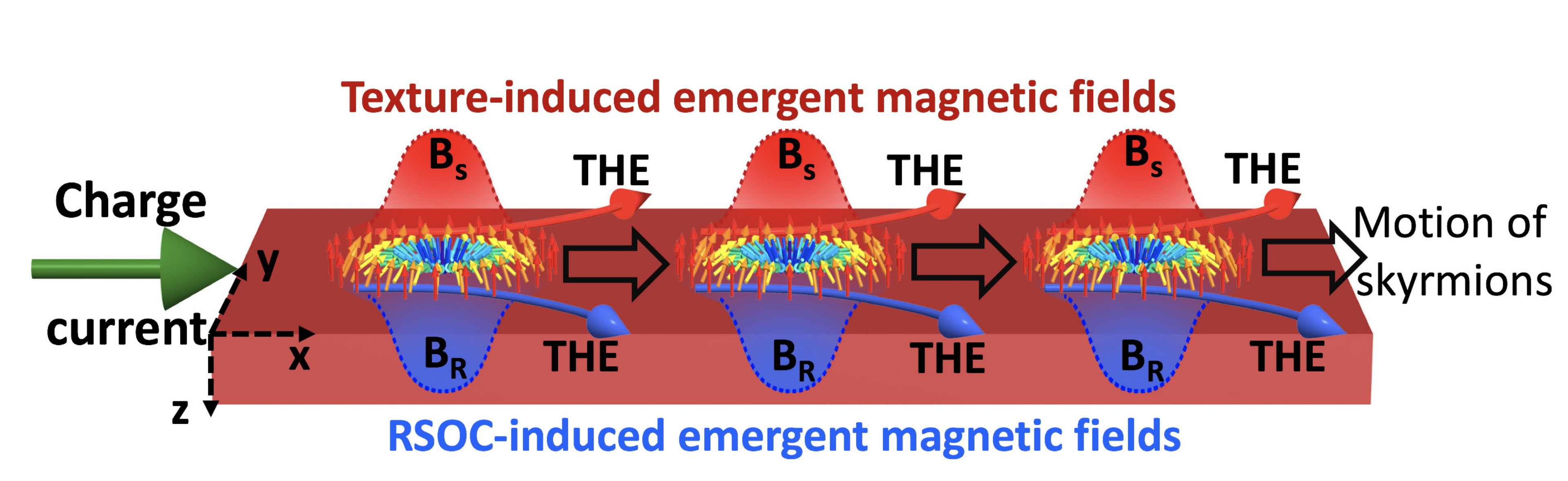}}
\caption{(Color online) Schematic illustration of the current-driven motion of \textit{N\'eel-type skyrmions}  array in the presence of SOC. Fields ${\bm B}_s$ (red) and  ${\bm B}_R$ (blue)  act in opposite directions, leading to the topological Hall effect (THE) on traversing electrons also in opposite directions.  As such, tuning ${\bm B}_R$ through the strength of the SOC can produce 
current-driven motion without skyrmion Hall effect (black arrows).}
\label{fig:Figure1}
\end{figure*}
It is well established that when itinerant electrons traverse a smooth and slowly varying magnetic texture, ${\bm m}({\bm r},t)$, there is a reorientation of their spins to follow the direction of local magnetization. This process gives raise to \textit{fictitious} electromagnetic fields that act on the itinerant electrons. The emergent electrodynamics resulting from the system described by Eq.~(\ref{eq:ham0}) is derived following the standard approach, i.e., the exchange term is diagonalized via a unitary transformation $\hat{U} = {\bm n}\cdot \hat{\bm \sigma}$, where ${\bm n} = \big(\cos\Phi\sin\theta/2, \sin\Phi\sin\theta/2 , \cos\theta/2 \big)$ 
in the spin-space. \cite{Tatara1994a,Tatara1994b} In the adiabatic limit, itinerant electrons in the transformed frame are subjected to a uniform ferromagnetic state and weakly coupled to the \textit{spin gauge fields} with the two contributions: (i) due to the magnetic texture ${\bm A}_s^\sigma$, and (ii) as a result of an interplay between SOC and the magnetic texture ${\bm A}_{ so}^\sigma = {\bm A}_{ R}^\sigma  + {\bm A}_{ D}^\sigma $, where ${\bm A}_{ R(D)}^\sigma$  is the RSOC (DSOC)-induced spin gauge field  given by \cite{Akosa2018b} 
\begin{subequations}\label{eq:sqfs}
\begin{align}
{\bf A}_s^\sigma &= -\frac{\sigma\hbar}{2e} \big([{\bm m} \times\partial_i {\bm m}]\cdot{\bm m}\big) {\bm e}_i, \\
{\bf A}_{ R}^\sigma &= \frac{\sigma\hbar}{2e}\frac{\big(m_x {\bm e}_y - m_y {\bm e}_x\big)}{\lambda_{ R}},  \\
{\bf A}_{ D}^\sigma &= -\frac{\sigma\hbar}{2e} \frac{\big(m_x {\bm e}_x - m_y {\bm e}_y\big)}{\lambda_{ D}},
\end{align}
\end{subequations}
where $i = x,y$, and $\lambda_{ R(D)} = \hbar^2/2m^*\beta_{ R(D)}$ is the characteristic length scale of the RSOC (DSOC). \cite{Kim2013} While the previous studies have focused on the effect of the SOC-induced emergent electric field on the itinerant electrons (spin-motive force), here, we focus on the effect of the corresponding emergent magnetic field on the itinerant electrons (Lorentz force). In the strong exchange limit, these fields are calculated from the corresponding spin gauge fields using  
\begin{eqnarray}
{\bm B}_{\eta}^\sigma = {\bm \nabla}\times {\bm A}_{\eta}^\sigma = \sigma {\rm B}_{\eta} {\bf e}_z ,
\end{eqnarray}
where $\eta = s, R, D$, and the field components are
\begin{subequations}\label{eq:emb}
\begin{eqnarray}
{\rm B}_s &=& -\frac{\hbar}{2e} \big([\partial_x {\bf m}\times \partial_y{\bf m}] \cdot {\bf m}  \big), \\
{\rm B}_{ R} &=& \frac{\hbar}{2e} \frac{\big(\partial_xm_x + \partial_y m_y \big)}{\lambda_{ R}} + \mathcal{O}(\beta_R^2),  \\
{\rm B}_{ D} & =& \frac{\hbar}{2e} \frac{\big(\partial_xm_y + \partial_y m_x \big) }{\lambda_{ D}}+ \mathcal{O}(\beta_D^2).
\end{eqnarray}
\end{subequations}
It turns out that ${\bm B}_s$ and ${\bm B}_R$ in Eq.~(\ref{eq:emb}) act in the opposite directions for N\'eel-type skyrmions as illustrated in Fig.~\ref{fig:Figure1}. 
The direct consequence of this fact is that the electrons traversing an array of N\'eel-type skyrmions in the positive {\bf x}-direction, experience two opposite fictitious magnetic fields: 
${\bm B}_s$ in the positive {\bf z}-direction that tends to deflect the electrons in the positive {\bf y}-direction and ${\bm B}_R$ in the negative {\bf z}-direction that tends to 
deflect the electrons in the negative {\bf y}-direction.  Since ${\bm B}_R$ has two free parameters $r_s$ and $\lambda_R$, by tuning 
them, one can realize a condition when ${\bm B}_R$ completely cancels out ${\bm B}_s$. In this case, the transverse \textit{Lorentz force} acting on electrons transversing the skyrmions is completely suppressed, or equivalently the \textit{Magnus force} on the magnetic structure is vanishing.

\begin{figure}[t!]
\centerline{\includegraphics[width=8.0cm]{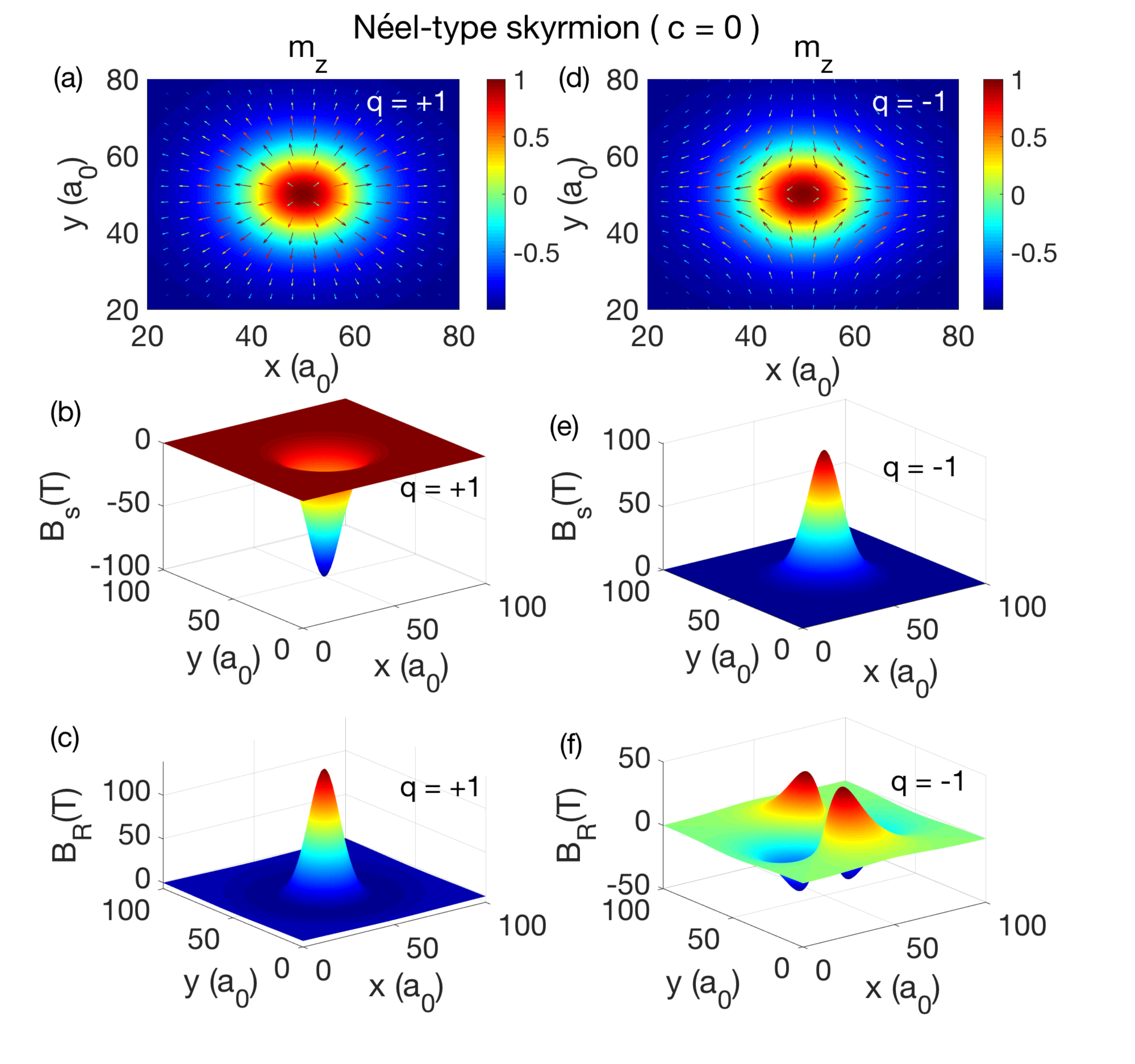}}
\caption{(Color online) Schematic diagram of the spatial magnetization profile for a N\'eel-type skyrmion (a) and antiskyrmion (b), and their corresponding emergent magnetic 
fields (c, e) and (d, f), respectively, in the presence of RSOC.}\label{fig:fig_1a}
\end{figure}

To gain more insight into the physics originating from the interplay of the different contributions to emergent magnetic fields given by Eq.~(\ref{eq:emb}), we consider a discrete 
square system of size $101\times101 a_0^2$, with the equilibrium skyrmion radius $r_s = 12 a_0$, where $a_0 = 0.3 \mbox{ nm}$ is the lattice constant. Furthermore, we take the 
effective mass of electrons $m^* = 0.4m_0$, where $m_0$ is the bare mass of electrons, and the RSOC strength $\beta_{R} = 2.5\times 10^{-11} \mbox{eV} \mbox{ m}$ 
(equivalent to $\lambda_{R} = 3.8 \mbox{ nm}$). Then we calculate the corresponding magnetic fields for different vorticities and helicities. In Fig.~\ref{fig:fig_1a}, the results for \textit{N\'eel-type skyrmions} $(q=+1)$ and \textit{antiskyrmions} $(q=-1)$ are shown with the magnetization profiles given by those presented in Figs.~\ref{fig:fig_1a}\textcolor{blue}{(a)} and \textcolor{blue}{(d)}, respectively. 
We see that for \textit{N\'eel-type skyrmions}, indeed, the fictitious magnetic fields ${\bm B}_s$ [c.f. Fig.~\ref{fig:fig_1a}\textcolor{blue}{(b)}] and ${\bm B}_R$ 
[c.f. Fig.~\ref{fig:fig_1a}\textcolor{blue}{(c)}] act in opposite directions such that, it is possible to achieve a current-driven motion without SkHE by tuning the strength of the RSOC. However, in the case of a \textit{N\'eel antiskyrmion} as shown in Figs.~\ref{fig:fig_1a}\textcolor{blue}{(e)} and 
\textcolor{blue}{(f)}, even though the transversing electrons experience these fictitious magnetic fields, the effect of ${\bm B}_R$ on its trajectory cancels out by symmetry 
[c.f. Fig.~\ref{fig:fig_1a}\textcolor{blue}{(f)}]. As such, only ${\bm B}_s$ influences its trajectory leading to SkHE for current-driven motion.

\begin{figure}[t!]
\centerline{\includegraphics[width=8.0cm]{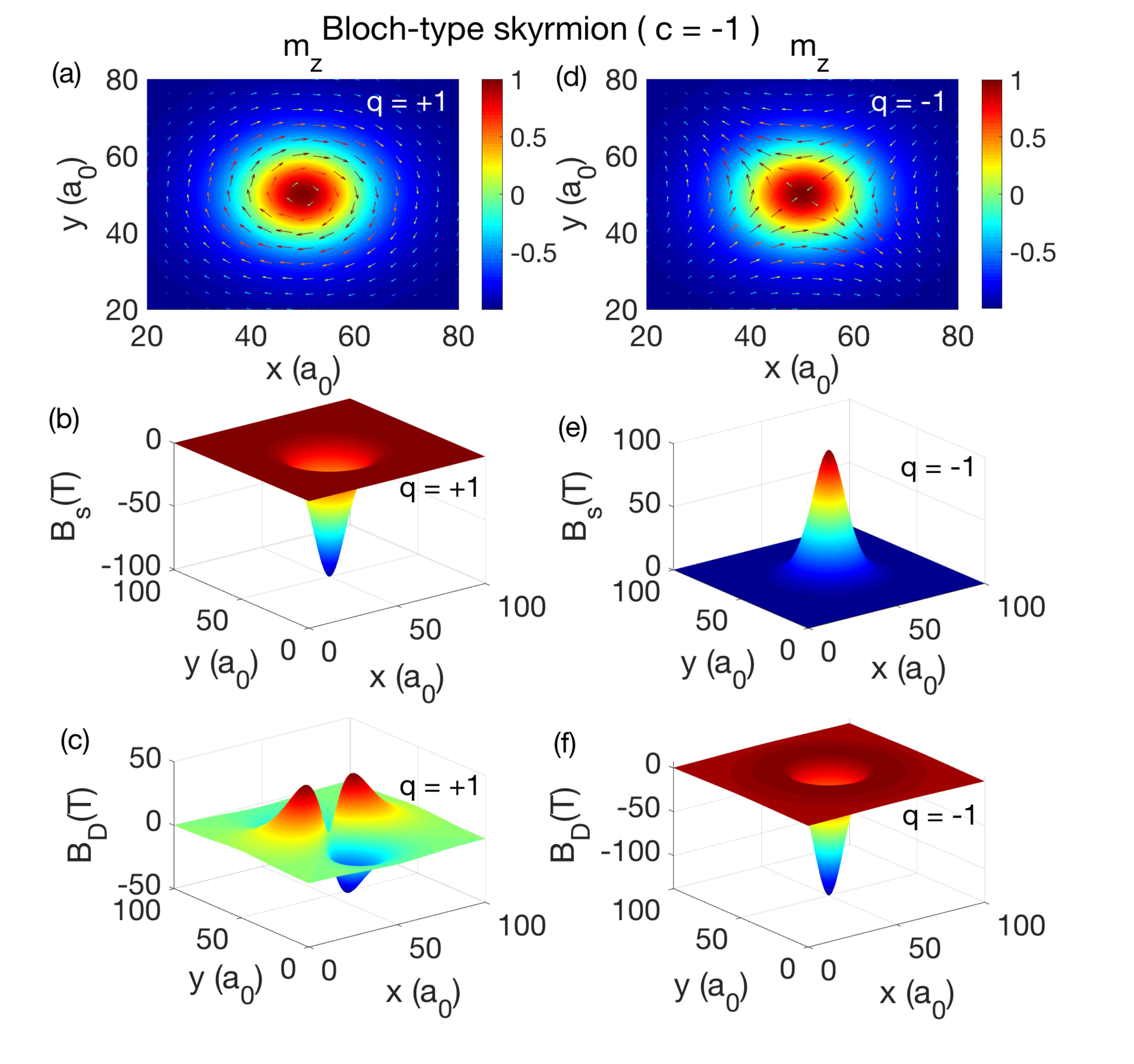}}
\caption{(Color online) Schematic diagram of the spatial magnetization profile for a Bloch-type skyrmion (a) and antiskyrmion (b), and their corresponding emergent magnetic 
fields (c, e) and (d, f), respectively, in the presence of DSOC.}
\label{fig:fig_1b}
\end{figure}

Similar arguments can be used to explain the characteristics of Bloch-type \textit{skyrmions} and \textit{antiskyrmions} as shown in Fig.~\ref{fig:fig_1b}, with the magnetization profiles as 
depicted in Figs.~\ref{fig:fig_1b}\textcolor{blue}{(a)} and \textcolor{blue}{(d)}, respectively. It turns out that unlike in the N\'eel-type case, the SOC-induced fictitious 
magnetic field ${\bm B}_D$, does influence the trajectory of electrons traversing \textit{Bloch-type skyrmions}, since the latter cancels out by symmetry 
[c.f. Fig.~\ref{fig:fig_1b}\textcolor{blue}{(c)}]. As such the trajectory of traversing electrons are detected by ${\bm B}_s$ [c.f. Fig.~\ref{fig:fig_1b}\textcolor{blue}{(b)}] leading to SkHE 
for current-driven motion. However, in the case of \textit{Bloch-type antiskyrmions}, ${\bm B}_s$ [c.f. Fig.~\ref{fig:fig_1b}\textcolor{blue}{(e)}] and ${\bm B}_D$
[c.f. Fig.~\ref{fig:fig_1b}\textcolor{blue}{(f)}] act in the opposite directions and as a result, by tuning the strength of DSOC and/or the $r_s$ via material engineering, it is possible to 
achieve a SkHE-free current-driven motion of Bloch antiskyrmions.

Therefore, our analysis shows that, depending on the vorticity and chirality of ferromagnetic solitons, it is possible to achieve a current-driven motion of the latter without SkHE via the engineering of the spin-orbit interaction in the system.  Although the heuristic analysis presented above, captures the important physics, In what follows, we provide a more rigorous argument based on the average emergent magnetic field that electrons traversing magnetic skyrmions experience. From Eq.~(\ref{eq:emb}), the average fictitious magnetic fields are calculated  as
\begin{subequations}\label{eq:amf}
\begin{eqnarray}\label{eq:avt}
\langle {\bm B}_s\rangle_{\rm av} &=&  -\frac{s\hbar}{e} \frac{2pq}{r_s^2} {\bm e}_z \\ \label{eq:avr}
\langle {\bm B}_R\rangle_{\rm av} &=&  \frac{s\hbar}{e} \frac{p(1 + q)}{2\lambda_R r_s} {\bm e}_z +  \boldsymbol{\mathcal{O}}(\beta_R^2) \\ \label{eq:avd}
\langle {\bm B}_D\rangle_{\rm av} &=&  \frac{s\hbar}{e} \frac{cp(1 - q)}{2\lambda_R r_s} {\bm e}_z +  \boldsymbol{\mathcal{O}}(\beta_D^2).
\end{eqnarray}
\end{subequations}
We immediately infer from Eq.~(\ref{eq:amf}) that up to the linear order in SOC, the average SOC-emergent magnetic field: (i) vanishes for N\'eel antiskyrmions 
[c.f. $q=-1$, in Eq.~(\ref{eq:avr})] and Bloch skyrmions [c.f. $q=+1$, in Eq.~(\ref{eq:avd})]  (ii) is finite and is opposite to ${\bm B}_s$ for N\'eel skyrmions 
[c.f. $q=+1$, in Eq.~(\ref{eq:avr})] and Bloch antiskyrmions [c.f. $q=-1$, in Eq.~(\ref{eq:avd})]. As a result, since the topological Hall effect of itinerant electrons as the 
traverse magnetic skyrmions is detected by these average fictitious magnetic fields, we recover the conclusions discussed in our heuristic analysis above.

Interestingly, these average additional fictitious magnetic fields are proportional to $ \beta_{R,D} /r_s$, and since the DMI has a subtle dependence on 
$r_s$, \cite{Kiselev2011, Rohart2013} but is proportional to the strength of the SOC \cite{Kikuchi2016,Koretsune2018}, one expects that the former should 
be at least dependent on the strength of the DMI in these systems. This theoretical prediction provides an interesting direction to explore to tune and potentially completely overcome 
this undesirable SkHE for spintronic applications. An interesting extension, which is however out of the scope of the present work, would be to directly investigate this effect via micromagnetic simulations. This is achievable via for example, taking into account the effect of the spin torques induced by the fictitious magnetic fields in Eq.~(\ref{eq:emb}). Indeed, previous studies have incorporated the textured-induced magnetic and electric fields and shown that this gives raise to the so-called topological torque and topological damping that directly influence the mobility of skyrmions. \cite{Bisig2016,Akosa2017}

\begin{figure}[t!]
\centerline{\includegraphics[width=6.0cm]{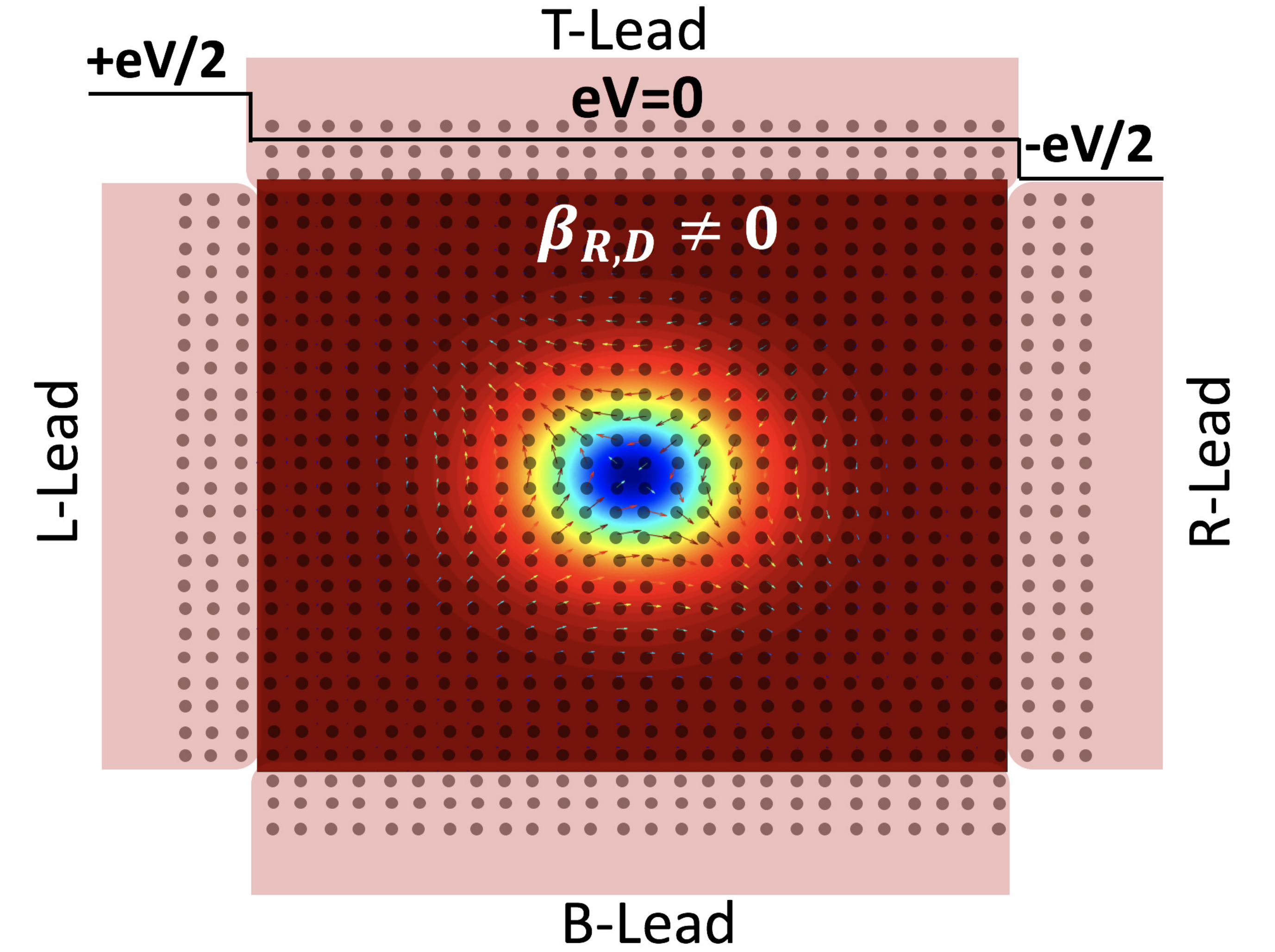}}
\caption{(Color online) Schematic diagram of four-terminal setup made up of a central region containing a magnetic skyrmion in the presence of SOC attached to four 
ferromagnetic leads L, T, R, B. The system is subjected to a longitudinal voltage bias of $eV$ while the transverse leads measure the Hall respond of the system.}
\label{fig:fourt}
\end{figure}

\section{Numerical Results}
We corroborate our analytical predictions via numerical calculations of the THE of electrons as they traverse an isolated skyrmion in the presence of RSOC and DSOC.  
Our considerations are based on a two-dimensional tight-binding model on a square lattice, described by the Hamiltonian 
\begin{equation}\label{eq:tbh}
\mathcal{H}= \sum_{\bm i} \hat{c}_{\bm i}^\dagger \big( \epsilon_{\bm i}  + J {\bm m}_{\bm i}\cdot \hat{\bm \sigma} \big)\hat{c}_{\bm i} -  
\sum_{<{\bm i},{\bm j}>}\hat{c}_{\bm i}^\dagger t_{{\bm i} {\bm j}} \hat{c}_{\bm j} ,
\end{equation}
where  $\epsilon_{\bm i}$ and $\hat{c}_{\bm i}^\dagger (\hat{c}_{\bm i})$ are the onsite energy and the spinor creation (annihilation) operators on site ${\bm i} = ({i_x,i_y})$, 
respectively. $J$ is the exchange energy that couples the spin of electrons $\hat{\bm \sigma}$ to local magnetization ${\bm m}_{\bm i}$, and $t_{{\bm i} {\bm j}} $ is 
the nearest-neighbor hopping that incorporates the spin-orbit interaction and is given by
\begin{equation}
t_{{\bm i} {\bm j}} = \left \{
  \begin{aligned}
 & t_0  + i t_{R} \sigma_y + i t_{D} \sigma_x , &&  \ {\bm j} = {\bm i} \pm {\bm x},\\
 & t_0  - i t_{R} \sigma_x - i t_{D} \sigma_y, && \  {\bm j} = {\bm i} \pm {\bm y} ,
  \end{aligned} \right.
\end{equation}
Here $t_0$ is the hopping in the absence of SOC, $t_{ R(D)} = \beta_{R(D)}/2a_0$ and $a_0$ is the lattice constant. We note that at low band filling, there is direct correspondence 
between the continuous and discrete Hamiltonians given in Eq.~(\ref{eq:ham0}) and Eq.~(\ref{eq:tbh}), respectively for $t_0 = \hbar^2/2m^*a_0^2$ \cite{Nikolic2005}. An isolated 
skyrmion of radius $r_s = 10a_0$ in embedded in a ferromagnetic background to which four ferromagnetic leads are attached as depicted in Fig.~\ref{fig:fourt}. 
We employ the Landauer-B\"uttiker formalism \cite{Buttiker1986} to investigate the coherent charge transport in our system which we subject to a longitudinal bias voltage across the 
left (L) and right (R) leads and measure the transverse responds via the top (T) and bottom (B) leads. The terminal current of the $\mu$-lead is calculated as
\begin{equation}
I_\mu = \frac{e^2}{2\pi\hbar} \sum_{\mu\ne\nu}(T_{\nu \mu}V_\mu - T_{\mu \nu}V_\nu), 
\end{equation}
where $V_\mu$ is the voltage at the $\mu$-lead and $T_{\nu \mu}$ is the transmission coefficient for electrons from the $\mu$-lead to the $\nu$-lead, which is 
calculated via the use of the KWANT software package \cite{Groth2014}. The terminal voltages are calculated following Ref.~\onlinecite{Ndiaye2017}, from which the 
THE is quantified via the topological Hall angle defined as 
\begin{equation}
\theta_{\rm TH} = \frac{V_T - V_B}{V_R - V_L}.
\end{equation}
We consider the strong exchange limit with $J = 5t_0$, and investigate the dependence of $\theta_{\rm TH}$ on the strength of RSOC ($t_R$) for N\'eel-type and 
DSOC ($t_D$) for Bloch-type skyrmions and antiskyrmions. Furthermore, to rule out any possibility that our results stems from size-effect, we perform a systematic study 
with different system sizes $L\times L$, for $L = 101a_0$, $161a_0$, $181a_0$ and an optimal (to ensure smooth enough magnetization variation from system to 
leads \cite{Akosa2017}) skyrmion radius of $10a_0$.

\begin{figure*}[t!]
\centerline{
\begin{tabular}{lcc}
\includegraphics[width=6.0cm]{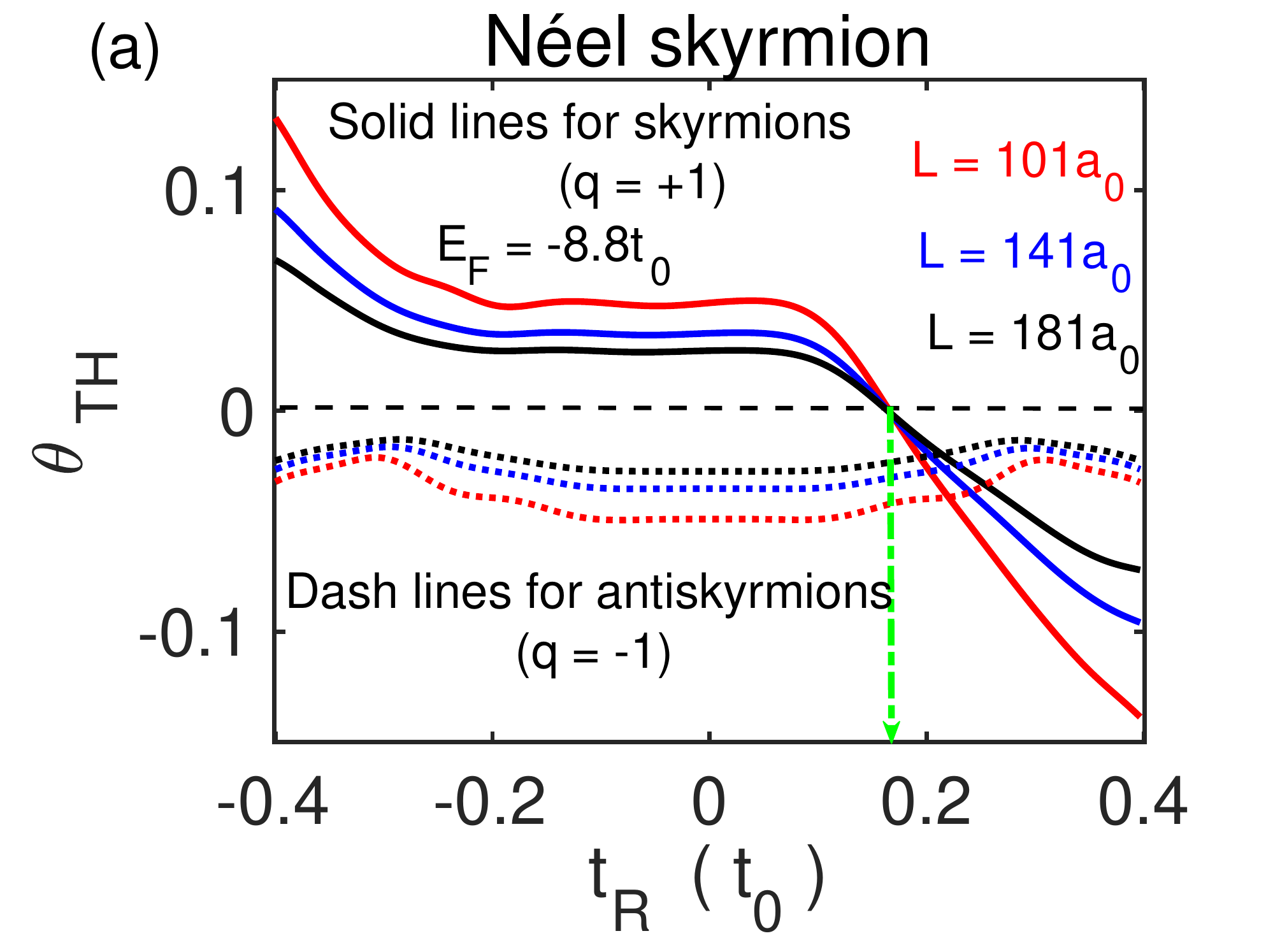} &
 \includegraphics[width=6.0cm]{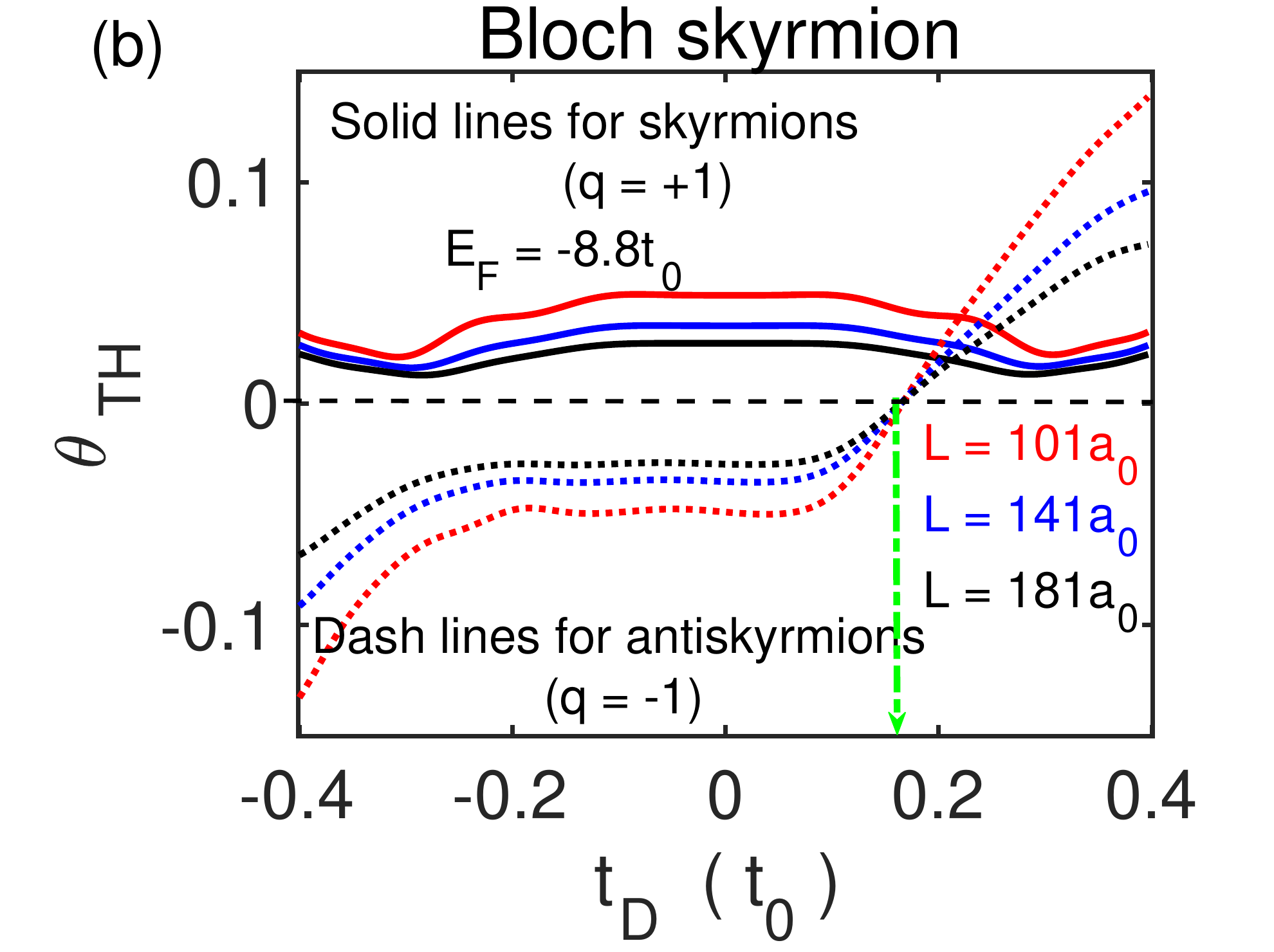} 
\end{tabular}}
\caption{(Color online) Dependence of the THE on the strength of (a) RSOC ($t_R$) for N\'eel skyrmion and (b) DSOC ($t_D$) for Bloch skyrmion, for different system sizes. 
Notice that the value of $t_R$ for N\'eel skyrmion and $t_D$ for Bloch antiskyrmion at which the THE vanishes (green arrow) is independent on the system size.}\label{fig:num}
\end{figure*}

Our numerical results as shown in Fig.~\ref{fig:num}, confirm the physics underscored by our analytical derivations.  Furthermore, the green arrow in 
Figs.~\ref{fig:num} \textcolor{blue}{(a)} and \textcolor{blue}{(b)} shows that the value of $t_R$ for N\'eel skyrmion and $t_D$ for Bloch antiskyrmion at which the THE vanishes
(green arrow) is independent on the system size as such, we rule out the possibilities that the observe results are artifact from size effect.

\section{Conclusions}
Magnetic skyrmions are hugely considered as a contender for information carriers in future spintronic applications. However, the parasitic SkHE constitutes a 
technological challenge for the integration of the former in such applications. Several theoretical proposals, which focus on suppressing the Magnus force that 
gives raise to this detrimental SkHE, have been put forward.  In this study, we focus on exploring the possibilities of overcoming the SkHE via tuning spin-orbit interactions that are inherent in skyrmionic systems. Starting from the emergent electrodynamics in the latter in the presence of SOC, we demonstrate that the additional SOC-induced emergent fictitious magnetic fields can be used to tune the SkHE. Our calculations show that by tuning the strength of RSOC in N\'eel skyrmions or DSOC in Bloch 
antiskyrmions, it is possible to achieve a current-driven motion without SkHE in these systems. Our findings open up promising perspective on overcoming the SkHE.

This work was supported by Grant-in-Aid for Exploratory Research (No.~16K13853) and Grant-in-Aid for Scientific Research (B) (No.~17H02929) from Japan Society for the 
Promotion of Science (JSPS). H.\,L. acknowledges support from HeNan University (Grant No.~CJ3050A0240050) and National Natural Science Foundation of China 
(Grants No.~11804078). O.\,A.\,T. acknowledges support by the Cooperative Research Project Program at the Research Institute of Electrical Communication, Tohoku University and by Ministry of Science and Higher Education of the Russian Federation in the framework of Increase Competitiveness Program of NUST "MISiS" (No.\,K2-2019-006), implemented by a governmental decree dated 16th of March 2013, N 211. C. A. A. thanks A. Abbout for useful discussions.

\end{document}